\documentclass[aip,
 amsmath,amssymb,
 reprint,%
]{revtex4-1}

\usepackage{graphicx}
\usepackage{color}
\usepackage{bm}
\usepackage{amsmath}
\usepackage{amssymb}
\usepackage{hyperref}
\usepackage{float}

\def\xcm3{\mbox{cm}^{-3}}

\def\Wxcm2{\mbox{Wcm}^{-2}}

\def\Axm2{\mbox{Am}^{-2}}

\definecolor{red}{rgb}{1,0,0}
\definecolor{blue}{rgb}{0,0,1}

\begin{document}

\author{M. E. Dieckmann}\email[]{mark.e.dieckmann@liu.se}
\affiliation{Department of Science and Technology, Link\"oping University, SE-60174 Norrk\"oping, Sweden}\affiliation{Kavli Institute for Theoretical Physics, University of California at Santa Barbara, CA 93106, USA}

\author{J. D. Riordan}
\affiliation{Department of Physics, University College Cork, Cork T12 K8AF, Ireland}

\author{A. Pe'er}
\affiliation{Department of Physics, Bar-Ilan University, Ramat-Gan, 5290002 Israel}

\date{\today}

\title{Change of a Weibel-type to an Alfv\'enic shock in pair plasma by upstream waves}

\begin{abstract}
We examine with particle-in-cell (PIC) simulations how a parallel shock in pair plasma reacts to upstream waves, which are driven by escaping downstream particles. Initially, the shock is sustained in the two-dimensional simulation by a magnetic filamentation (beam-Weibel) instability. Escaping particles drive an electrostatic beam instability upstream. Modifications of the upstream plasma by these waves hardly affect the shock. In time, a decreasing density and increasing temperature of the escaping particles quench the beam instability. A larger thermal energy along than perpendicular to the magnetic field destabilizes the pair-Alfv\'en mode. In the rest frame of the upstream plasma, the group velocity of the growing pair-Alfv\'en waves is below that of the shock and the latter catches up with the waves. Accumulating pair-Alfv\'en waves gradually change the shock in the two-dimensional simulation from a Weibel-type shock into an Alfv\'enic shock with a Mach number that is about 6 for our initial conditions.
\end{abstract}


\maketitle

\section{Introduction}

Black hole microquasars are binary systems, in which a stellar-mass black hole accretes material from a nearby companion star and ejects a relativistic jet \cite{Margon1979,Falcke1996,Fender2004}. The interior of the jet is composed of electrons and positrons and an unknown fraction of ions. It is thought that the low number density and the high temperature of the jet material let the effects of binary collisions between particles be small compared to those of the electromagnetic fields, which are driven by the ensemble of all particles. We call such a plasma collisionless. 

A nonuniform flow speed of the jet can result in collisionless shocks at those locations where a faster flow is catching up with a slower one. A shock can also form between the relativistically moving jet material and the inner cocoon of the jet \cite{Bromberg2011}, which is separated by a collisionless magnetic discontinuity from the ambient plasma into which the jet expands \cite{Dieckmann2019}. If the jet is leptonic then we would expect that its internal shocks slow down and heat a flow of electrons and positrons. 

Shocks in collisionless plasma have a finite thickness. The shock transition layer is defined as the spatial interval where the inflowing upstream plasma is slowed down, heated and compressed by the electromagnetic fields that mediate the shock. The upstream material moves at a supersonic speed in the reference frame of the shock. Supersonic means in this context that the flow speed exceeds the speed of the wave that mediates the shock. Plasma, which crossed the transition layer, enters the downstream region. The downstream material is a hot and dense plasma in a thermal equilibrium that moves at a subsonic speed relative to the shock. 

Collisionless leptonic shocks have been studied widely in the past with particle-in-cell (PIC) simulations. Relativistic collision speeds and a low temperature of the upstream plasma yield shock transition layers that are mediated by the filamentation instability, which is also known as the beam-Weibel instability \cite{Kazimura1998,Spitkovsky2005,Chang2008,Nishikawa2009,Sironi2009,Marcowith2016}. The wavevectors \textbf{k} of these magnetowaves are oriented primarily orthogonally to the flow direction of the upstream plasma. The magnetowaves heat the upstream plasma, which crosses the transition layer, to a relativistic temperature before it enters the downstream region.

The exponential growth rate of the filamentation instability decreases as the collision speed is decreased. Oblique modes can outgrow the filamentation modes in particular if the interacting plasma beams have different densities. A nonrelativistic leptonic shock was examined in the simulation in Ref.\cite{Dieckmann2017}. Escaping energetic downstream particles interacted with the inflowing upstream plasma via an electrostatic oblique mode instability. The temperature of the preheated upstream plasma was higher along the collision direction than orthogonal to it after it crossed this layer. This thermal anisotropy resulted in a magneto-instability similar to the one found by Weibel\cite{Weibel1959}, which thermalized the pair plasma and established its thermal equilibrium. Increasing the shock speed to a mildly relativistic value triggered a filamentation instability between the intervals where electrostatic waves and the Weibel modes grew \cite{Dieckmann2018MNRAS}. 

Collisionless shocks in pair plasma, which is not permeated by a background magnetic field $\mathbf{B}_0$, are mediated by waves driven by the two-stream instability, the oblique mode instability and the filamentation instability \cite{Bret2010}. Aligning $\mathbf{B}_0$ with the shock normal modifies the spectrum of unstable waves \cite{Hededal2005} and only the two-stream instability operates if $B_0=|\mathbf{B}_0|$ is sufficiently strong \cite{Bret2006}. Such a magnetic field does not only modify the dispersion relation of the aforementioned modes. It can also introduce new unstable wave modes. 

Consider for example a pair shock with a normal that is aligned with $\mathbf{B}_0$. This field can maintain different temperatures perpendicularly and parallel to $\mathbf{B}_0$. If the electromagnetic fields in the original shock transition layer cannot establish a thermally isotropic distribution of the plasma then instabilities like the mirror- or firehose instabilities \cite{Schlickeiser2010,Bret2018,DieckmannPPCF2019} can grow behind the original transition layer and broaden it. Another example is provided by shocks in magnetized electron-ion plasma. Such shocks can emanate beams of particles with a super-Alfv\'enic speed into their upstream regions. Such beams can trigger the growth of Alfv\'enic waves upstream of the shock, thereby broadening its transition layer. The cosmic-ray driven Bell instability \cite{Bell2004,Park2015,Caprioli2018} falls into this category. 

We study here with one- (1D) and two-dimensional (2D) PIC simulations the instabilities that grow in the transition layer of a shock in a pure pair plasma. A magnetic field is aligned with the shock normal. Its amplitude is not large enough to suppress the filamentation instability. The shock is created when the pair plasma, which is reflected at one boundary of the simulation box, interacts with the inflowing pair plasma. The two-stream instability grows first in the 1D simulation, which excludes geometrically the filamentation instability. It creates a plasma close to the reflecting boundary that is thermally anisotropic. Eventually a magneto-instability is triggered, which results in the growth of pair-Alfv\'en waves. In what follows, we refer to this instability as the pair-Alfv\'en wave instability. Pair-Alfv\'en waves mediate the shock in the 1D simulation. The filamentation instability outgrows two-stream instability in the 2D simulation and its magneto-waves sustain initially the shock. Some electrons and positrons outrun the shocks in both simulations. Initially these particles drive electrostatic instabilities upstream of the shock\cite{Dieckmann2018MNRAS}. In time pair-Alfv\'en waves \cite{Keppens2019} grow in the upstream region of the shocks. The shock catches up with these slow waves and piles them up. Pair-Alfv\'en waves partially replace the filamentation modes in the 2D simulation as the means to sustain the shock. A similar replacement of the filamentation mode by Alfv\'en waves driven by Bell's instability has been observed at quasi-parallel electron-ion shocks\cite{Crumley2019}.

This work is the first, to our knowledge, to identify in simulation results, and to extensively study these pair-Alfv\'en-mediated parallel shocks and their associated upstream turbulence in plasmas with significant background magnetic fields. This work is therefore significant in bettering our understanding of trans-relativistic, non-collisional, pair-plasma environments, such as those predicted to occur at the base of relativistic jets in black hole microquasars.

Our paper is structured as follows. Section 2 summarizes the PIC method and it lists our initial conditions. It also gives a brief summary of the instabilities that develop in our simulation. Section 3 presents the simulation results, which are discussed in Section 4.

\section{The PIC code and the initial conditions}

Amp\`ere's law $\mu_0 \epsilon_0 \dot{\mathbf{E}}=\nabla \times \mathbf{B}-\mu_0 \mathbf{J}$ and Faraday's law $\dot{\mathbf{B}}=-\nabla \times \mathbf{E}$ are approximated on a numerical grid, where $\mathbf{E}$, $\mathbf{B}$ and $\mathbf{J}$ are the electric field, the magnetic field and the current density. The vacuum permittivity and permeability are $\epsilon_0$ and $\mu_0$. Each plasma species $j$ is approximated by computational particles (CP's). The $i^{th}$ CP has a charge-to-mass ratio $q_i/m_i$ that must match that of the species $j$ it represents. The electromagnetic fields are coupled to the CPs and the CPs are coupled to $\mathbf{J}$ via suitable numerical schemes as implemented in the EPOCH code we use \cite{Arber2015}. 

Our two-dimensional simulation resolves $x$ by $2\times 10^4$ grid cells and $y$ by $2000$ grid cells. Boundary conditions are reflective along $x$ and periodic along $y$. We model one electron species and one positron species, which are uniformly distributed in space. Each species has the density $n_0/2$ and is resolved by a total of $8 \times 10^8$ CPs, which corresponds to 20 CPs per cell for each species. We perform also a one-dimensional simulation with the same plasma parameters, where we resolve only $x$ and use $10^7$ CPs for each species. This number amounts to 500 CPs per cell for electrons and the same number for the positrons. We do not inject CPs while the simulation is running.

The plasma frequency is $\omega_p = {(e^2n_0/m_e\epsilon_0)}^{1/2}$ with $e,m_e$ being the elementary charge and the electron mass. We normalize the time $t$ to $\omega_p^{-1}$, space to the plasma skin depth $\lambda_s = c/\omega_p$ ($c:$ light speed) and densities to $n_0$. Frequencies $\omega$ are normalized to $\omega_p$ and wavenumbers $k_x$ in 1D or $(k_x,k_y)$ in 2D to $\lambda_s^{-1}$. The simulation box spans the intervals $0 \le x \le 2650$ and $0 \le y \le 265$. Both species have a Maxwellian velocity distribution with the temperature $T_0$ = 10 keV, which gives the thermal speed $v_t = 4.2 \times 10^7$ m/s with $v_t = {(k_BT_0/m_e)}^{1/2}$ ($k_B:$ Boltzmann constant), and the mean speed $v_b = -3v_t$ ($0.42c$) along $x$. The pair cloud moves with the mean speed $v_b$ to the boundary at $x=0$. Particles are reflected by this boundary and flow back to increasing $x$. A shock is triggered by the interaction of the reflected and inflowing pair plasma. The Debye length is $\lambda_D=0.14$. We set the grid cell size to $\lambda_D$ and hence we resolve one skin depth $\lambda_s$ by just over 7 grid cells. Electric fields \textbf{E} and magnetic fields \textbf{B} are normalized to $m_e c \omega_p / e$ and $m_e \omega_p / e$ respectively. A magnetic field $\mathbf{B}_0=(B_0,0,0)$ with $B_0=0.1$ is present at $t=0$. This value equals the normalized electron gyro-frequency $\omega_c = eB_0/m_e\omega_p$. Our pair plasma has a value $\beta = n_0 k_B T_0/(B_0^2/2\mu_0)=4$. The pair Alfv\'en speed $v_A = B_0 / {(\mu_0 n_0 m_e)}^{1/2}$ equals $0.7v_t$ ($0.1c$). Both simulations are stopped at $t_{sim}=2500$.

\begin{figure}
\includegraphics[width=\columnwidth]{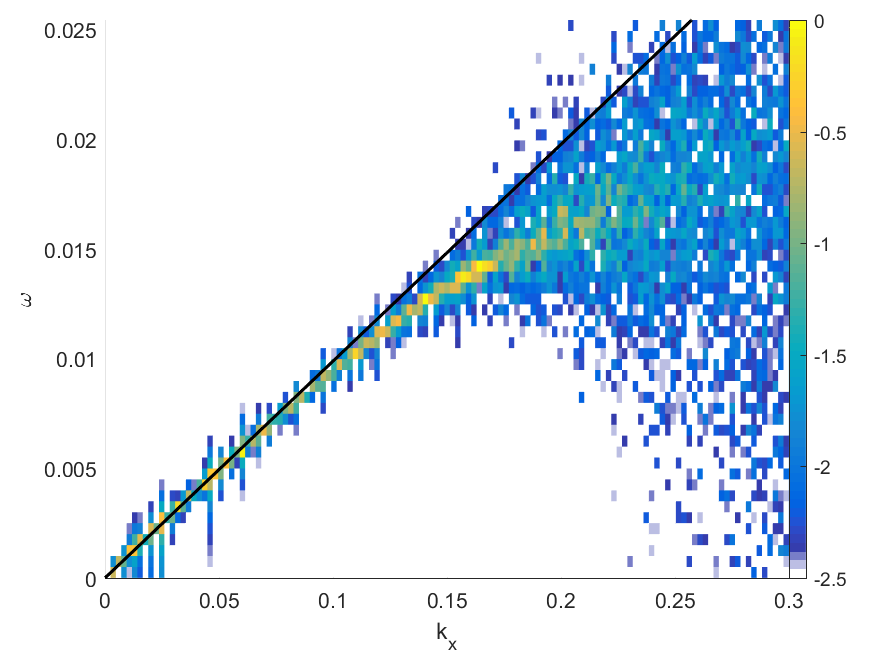}
\caption{Dispersion relation of the low-frequency electromagnetic waves in the simulation plasma. The color shows the power of $B_\perp^2(k_x,\omega)$ normalized to its peak value on a 10-logarithmic scale. The dispersion relation of the pair Alfv\'en mode in cold plasma (red curve) and $\omega = v_A k_x$ (black line) are overplotted. The cold plasma mode has its cyclotron resonance at $\omega = 0.1$.}
\label{figure1}
\end{figure}

Pair-Alfv\'en waves play a pivotal role in sustaining our shocks. Thermal noise provides us with insight into their dispersion relation. Most of its power is concentrated on the undamped modes in the simulation plasma \cite{Dieckmann2004}. Figure \ref{figure1} shows the frequency-wavenumber spectrum $B_\perp^2(k_x,\omega)=B_y(k_x,\omega)\bar{B}_y(k_x,\omega)+B_z(k_x,\omega)\bar{B}_z(k_x,\omega)$ for a pair plasma with the aforementioned plasma parameters in its rest frame (the bar denotes the complex conjugate). 
The thermal noise at low wave numbers peaks on the dispersion relation of the pair-Alfv\'en wave. Its phase speed is $v_A$ up to $k_x\approx 0.1$ (the wavelength is $\approx 60$). The solution of the linear dispersion relation in cold pair plasma\cite{Keppens2019} follows closely $\omega  = v_A k_x$ in this wavenumber interval and its phase speed hardly decreases for larger $k_x$ in the displayed interval. Its frequency converges to the cyclotron resonance $\omega_c = 0.1$ with increasing $k_x$. The phase speed of the mode in the simulation decreases faster than that of the cold plasma mode for larger values of $k_x$ and the sharp noise peak disappears for $k_x>0.25$, which usually implies that the wave is damped (See also Ref.\cite{DieckmannPPCF2019}). This damping is likely to be caused by wave-particle interactions because $v_A \approx 0.7v_t$.

Due to the presence of the background magnetic field, we have several wave modes in our simulation that can be rendered unstable by interacting pair beams and by a thermal anisotropy. These instabilities have been analysed under the assumption that the wave amplitudes are small and that the plasma can be approximated either by a bi-Maxwellian particle velocity distribution or by counter-streaming beams. 

We consider first the case of counterstreaming beams. The large initial speed modulus $|v_b|=3v_t$ implies that the inflowing and reflected pair plasmas form two beams close to the reflecting boundary that are separated in velocity space. Both beams flow along $\mathbf{B}_0$. It is of interest to determine whether or not our value $B_0=0.1$ is large enough to suppress the magnetic filamentation instability of counter-streaming beams. Bret et al.~\cite{Bret2006} determine the value of $B_0$ that is needed to suppress the filamentation instability of counter-streaming cold electron beams. For our non-relativistic flow speed and assuming that the inflowing and reflected pair clouds consist only of electrons and are equally dense, we obtain the critical magnetic field value $B_c=\sqrt{2}v_b/c$ giving $B_c=0.6$; we expect that our magnetic field is not strong enough to suppress the filamentation instability. This instability is, however, suppressed geometrically in our 1D simulation. We expect that the inflowing and reflected pair cloud trigger an electrostatic two-stream instability in the 1D simulation, which saturates by forming phase space holes \cite{Jao2014,Dieckmann2018MNRAS}. Individual phase space holes are stable in 1D but unstable otherwise\cite{Morse69}. Their collapse heats the plasma. As long as the simulation resolves more than one spatial dimension, two-stream instabilities can mediate a narrow shock transition layer. This shock transition layer widens in a 1D simulation\cite{Dieckmann2018MNRAS} because planar phase space holes can only thermalize by their slow coalescence\cite{Roberts67}.

Let us assume that the initial instabilities have heated up the plasma along the collision direction. Weibel considered in his work\cite{Weibel1959} a single electron species with a bi-Maxwellian non-relativistic velocity distribution. Electrons had a lower temperature along one direction than along the other two. He found aperiodically growing waves with a wave vector along the cool direction. He also showed that aligning $\mathbf{B}_0$ with this direction cannot stabilize the plasma. Weibel's work was extended to pair plasma\cite{Gary2009}. Aligning $\mathbf{B}_0$ with the cool direction of both species gives rise to two modes; Alfv\'en-like waves are positronic modes while the electrons give rise to magnetosonic-like waves. Both modes have an equal dispersion if electrons and positrons have equal distributions giving the combined mode a linear polarization \cite{Stewart1992}. We refer here to this mode as the pair-Alfv\'en mode. Gary et al.~\cite{Gary2009} consider first the case where the plasma temperature is the same in all directions. The pair-Alfv\'en mode is undamped in the wavenumber interval where the wave has the dispersion relation $\omega=v_Ak_x$, which is $k_x < 0.1$ in Fig.~\ref{figure1}. Increasing only the plasma temperature perpendicular to $\mathbf{B}_0$ gives rise to a mirror-like instability. 

Schlickeiser~\cite{Schlickeiser2010} examines also the case where the pair plasma is hotter along $\mathbf{B}_0$ than perpendicular to it, which is relevant for our simulation. The initial value $\beta = 4$ in our simulation is boosted along $\mathbf{B}_0$ because particles get reflected by the boundary at $x=0$ and mix with the inflowing plasma. Typical particle speeds along this direction are thus increased from the thermal speed $v_t$ to the beam speed modulus $|v_b|=3v_t$ in the simulation frame. If we assume that the thermal pressure perpendicular to the magnetic field remains unchanged, we obtain a thermal anisotropy $A=v_{t,\perp}^2 /v_{t,\parallel}^2 \ll 1$ ($v_{t,\perp}^2, v_{t,\parallel}$: thermal speeds perpendicular and parallel to $\mathbf{B}_0$). According to Fig. 8 in Ref.~\cite{Schlickeiser2010}, only the firehose instability can grow at low wavenumbers $k_x \ll 1$ if only wave vectors parallel to $\mathbf{B}_0$ are taken into account. It yields aperiodically growing fluctuations. The Fourier spectrum of a wave with an amplitude that grows exponentially and non-oscillatory involves waves with a wide frequency spread. Aperiodically growing waves can thus couple their energy into the low-frequency pair-Alfv\'en mode~\cite{DieckmannPPCF2019}.

\section{Simulation results}

\subsection{1D simulation}

Geometrical constraints suppress the filamentation instability in the 1D simulation because it grows by letting the plasma currents rearrange themselves in the direction orthogonal to the plasma collision direction. Pair-Alfv\'en waves and electrostatic Langmuir waves, which propagate along $\mathbf{B}_0 = (B_0,0,0)$, can still grow. 

\begin{figure*}
\includegraphics[width=\textwidth]{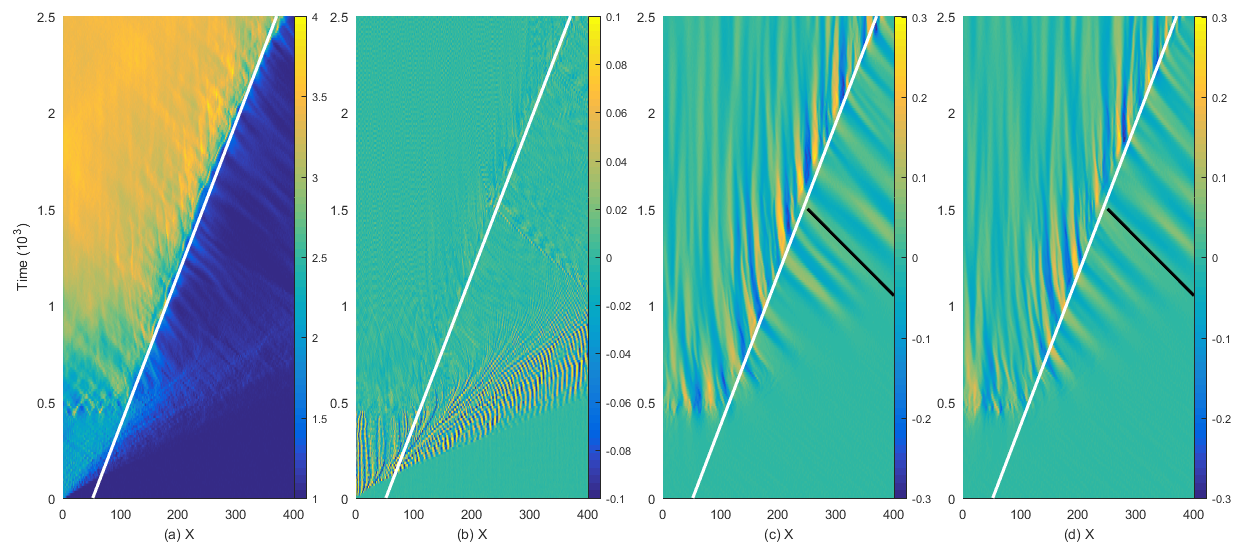}
\caption{Time evolution of the plasma in the 1D simulation: Panel (a) shows the plasma density. Panel (b) depicts the electric field $E_x$. Panels (c, d) display the magnetic field components $B_y$ and $B_z$. The overplotted white and black lines mark the speeds $v_f =0.12c$ and $v_w = -0.33c$.}
\label{figure2}
\end{figure*}

Figure \ref{figure2} shows the time evolution of the plasma density, that of the electric field $E_x$ and that of the magnetic field components $B_y$ and $B_z$. The $B_x$ component cannot change since $\nabla \cdot \mathbf{B}=0$. 
The plasma density close to $x=0$ increases to 2 after the simulation begins, due to the superposition of the inflowing and reflected plasma. We observe at early times density modulations, which are tied to oscillations of $E_x$ in Fig. \ref{figure2}(b). These waves are the result of a two-stream instability between the incoming and the reflected particles and they are present until $t=500$ in an interval that expands with time. 

Magnetowaves grow in Figs. \ref{figure2}(c, d) after the electrostatic waves collapse in the interval $0 \le x \le 150$. Both magnetic field components show wave activity in the interval $x \le 100$ during $500 \le t \le 1000$. Figure \ref{figure2}(a) shows that the plasma density close to the boundary increases from $2$ to $3.5$ during this time. The front of the dense plasma expands at the speed $v_f=0.12 c$ and it is correlated with strong waves. Some electrostatic waves propagate at a much larger speed. 

Figures \ref{figure2}(c, d) reveal waves that are propagating from the upstream
direction towards the front of the dense plasma after $t=500$. These waves are
transported with the upstream flow towards the dense plasma, then enter it
and finally are damped out. The wave amplitude is almost stationary to the left of the white line, where the plasma has a low mean speed. 

The frequency of the waves to the right of the white line is low in the rest
frame of the upstream plasma.  However, pair-Alfv\'en waves are the only
low-frequency waves that can propagate along $\mathbf{B}_0$ in the thermally
isotropic upstream plasma \cite{Keppens2019}. Their wavelength $2\pi/k_x\approx
50$ falls into the wavenumber interval where we find undamped pair-Alfv\'en
waves in Fig. \ref{figure1}. The phase speed of the magnetowaves is $-0.33c$ in
the box frame and $0.09c$ in the rest frame of the upstream plasma, which has a mean speed $-0.42c$. The waves we observe to the right of the white line in Figs. \ref{figure2}(c, d) are thus pair-Alfv\'en waves that propagate in the upstream direction. They are connected to the waves to the left of the white line, which suggests that they belong to the same wave branch.

\begin{figure}
\includegraphics[width=\columnwidth]{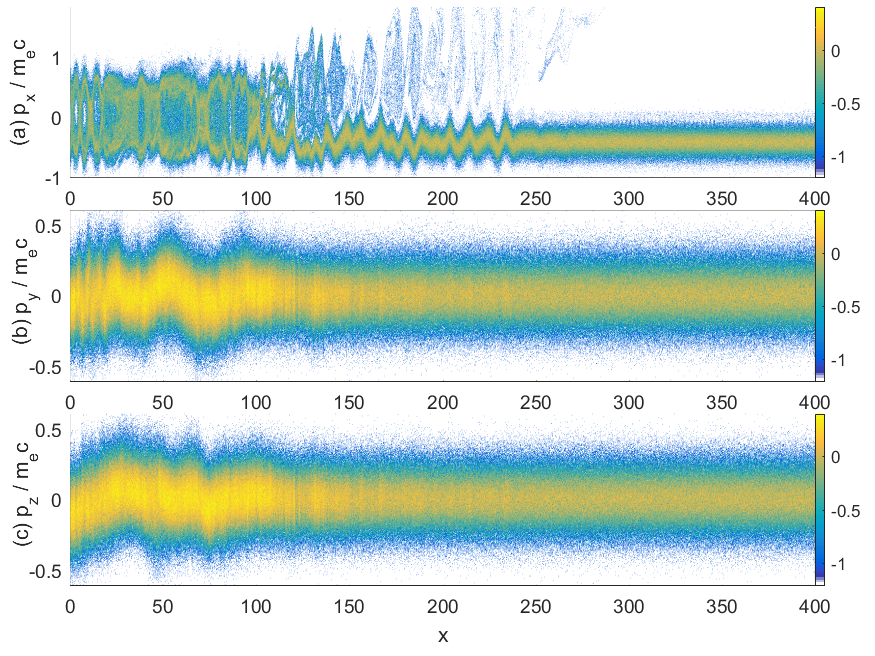}
\caption{Electron phase space density distribution at $t=380$: Panel (a) shows $f_e(x,p_x)$, panel (b) shows $f_e(x,p_y)$ and $f_e(x,p_z)$ is displayed by panel (c) where $p_{x,y,z}$ are the relativistic momenta in units of $m_ec$. All distributions are normalized to their peak value far upstream. A 10-logarithmic color scale is used.}
\label{figure3}
\end{figure}

The phase space density distribution sheds light on why the collapse of the electrostatic waves in Fig. \ref{figure2}(b) at $t\approx 500$ coincides with the growth of magnetowaves in Figs. \ref{figure2}(c, d). We select the electron distribution at the time $t=380$ and show its projections onto the three momentum axes in Fig. \ref{figure3}.
Figure \ref{figure3}(a) reveals phase space vortices in the interval $0 \le x \le 130$. They are the product of a two-stream instability between the inflowing and reflected electron beams. Dilute phase space vortices are present for $130 \le x \le 250$. They are responsible for the fast structures in Fig. \ref{figure2}(b). The other two projections show an increase of the phase space density for $0 \le x \le 150$ and some oscillations but the momentum range covered by the electrons is well below that in Fig. \ref{figure3}(a). The plasma in the overlap layer $x\le 130$ is hotter along $x$ than along the other directions. 

A pair-Alfv\'en wave instability lets the magnetic $B_y$ and $B_z$ components grow. These fields can deflect electrons and positrons from the parallel into the perpendicular direction and bring the plasma closer to thermal equilibrium; its density increases as observed in Figs \ref{figure2}(a). Pair-Alfv\'en waves are low-frequency waves, which explains why they do not grow immediately after the thermal anisotropy has developed. The pair plasma does not have a bi-Maxwellian distribution in the interval where the waves grow and the velocity distribution varies along $x$. The observed instability is thus not the firehose instability that was analyzed by Schlickeiser\cite{Schlickeiser2010}. Both instabilities may, however, be related.

\begin{figure}
\includegraphics[width=\columnwidth]{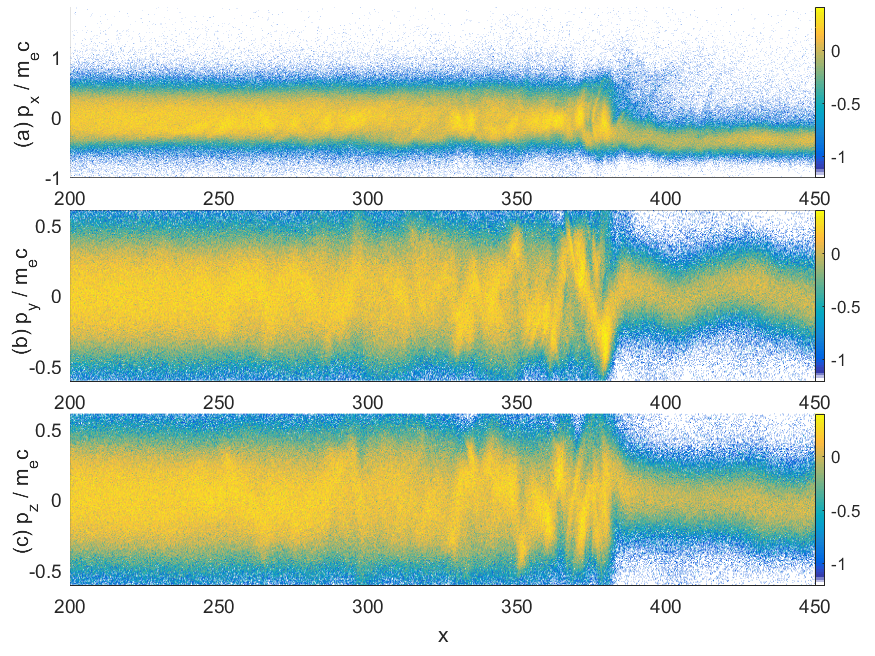}
\caption{Electron phase space density distribution at $t=2500$: Panel (a) shows $f_e(x,p_x)$, panel (b) shows $f_e(x,p_y)$ and $f_e(x,p_z)$ is displayed by panel (c). All distributions are normalized to their peak value far upstream. A 10-logarithmic color scale is used.}
\label{figure4}
\end{figure}

Figure \ref{figure4} displays the electron distribution close to the front of the dense plasma at the time $t_{sim}$. 
It reveals that this front is a shock, which is located at $x\approx 370$. It
thermalizes rapidly the inflowing upstream electrons as they cross a transition
layer in the interval $250 \le x \le 370$. Our simulation box is at rest in the
downstream frame of reference and $v_f=0.12c$ thus corresponds to the shock
speed in this frame. The phase speed of the pair-Alfv\'en wave in the downstream
plasma is reduced to $v_A^*=v_A/\sqrt{3.5}$ due to the higher plasma density
$\approx 3.5$. The shock moves at the speed $v_f \approx 2.3 v_A^*$.
Pair-Alfv\'en waves downstream of the shock can not keep up with this and so the
waves we observe must be generated directly behind the shock front by the
thermal anisotropy. The growth of these waves is accelerated by the seed waves,
which are transported with the upstream plasma to the shock. The lower Alfv\'en
speed and the higher plasma temperature downstream of the shock imply that the
waves can interact with the dense thermal bulk population of the particles,
explaining the damping of the pair-Alfv\'en waves in this region. The oscillations of the mean velocities along $y$ and $z$ in the region $x>370$ can be attributed to the pair-Alfv\'en waves that arrive at the shock from the upstream direction. 

The large velocity oscillations of the upstream plasma in Fig.~\ref{figure4} may couple the pair-Alfv\'en waves, which are polarized along $y$, with those that are polarized along $z$. Such a coupling would break the linear polarization these waves have when their amplitude is low.

\subsection{2D simulation}

\begin{figure*}
\includegraphics[width=\textwidth]{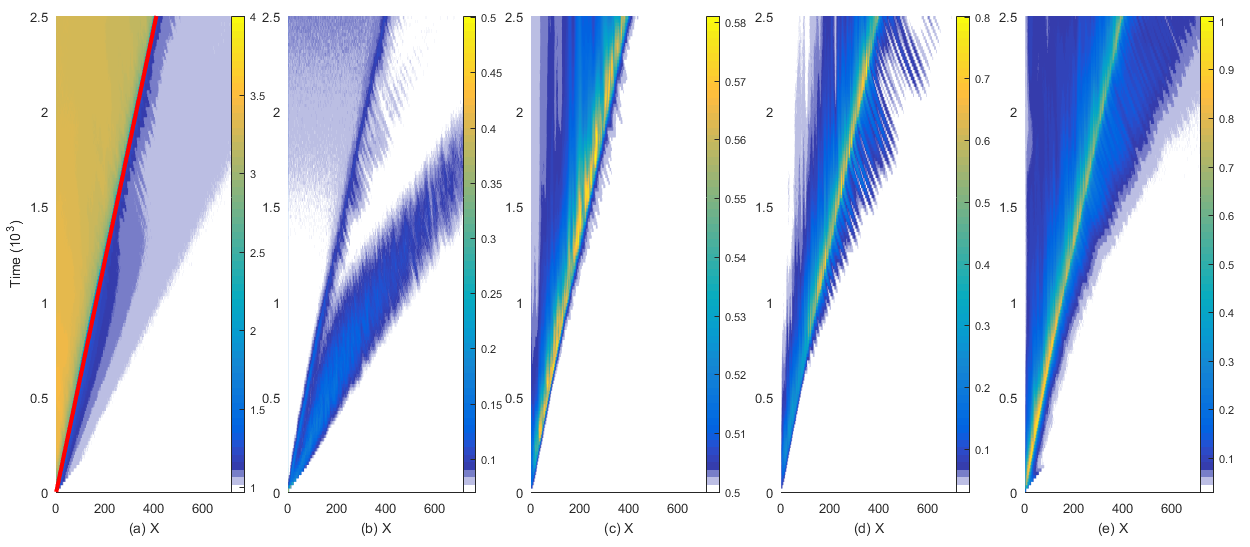}
\caption{Time evolution in the 2D simulation: Panel (a) shows the box-averaged plasma density. The red line marks the speed $v_{f2}=0.16c$. (b) displays the square root of the box-averaged normalized field energy $P_E$ of the in-plane electric field. Panels (c-e) show the square root of the box-averaged normalized field energies $P_{Bx}, P_{By}$ and $P_{Bz}$ of the magnetic $B_x$, $B_y$ and $B_z$ components, respectively.}
\label{figure5}
\end{figure*}

\begin{figure*}
\includegraphics[width=\textwidth]{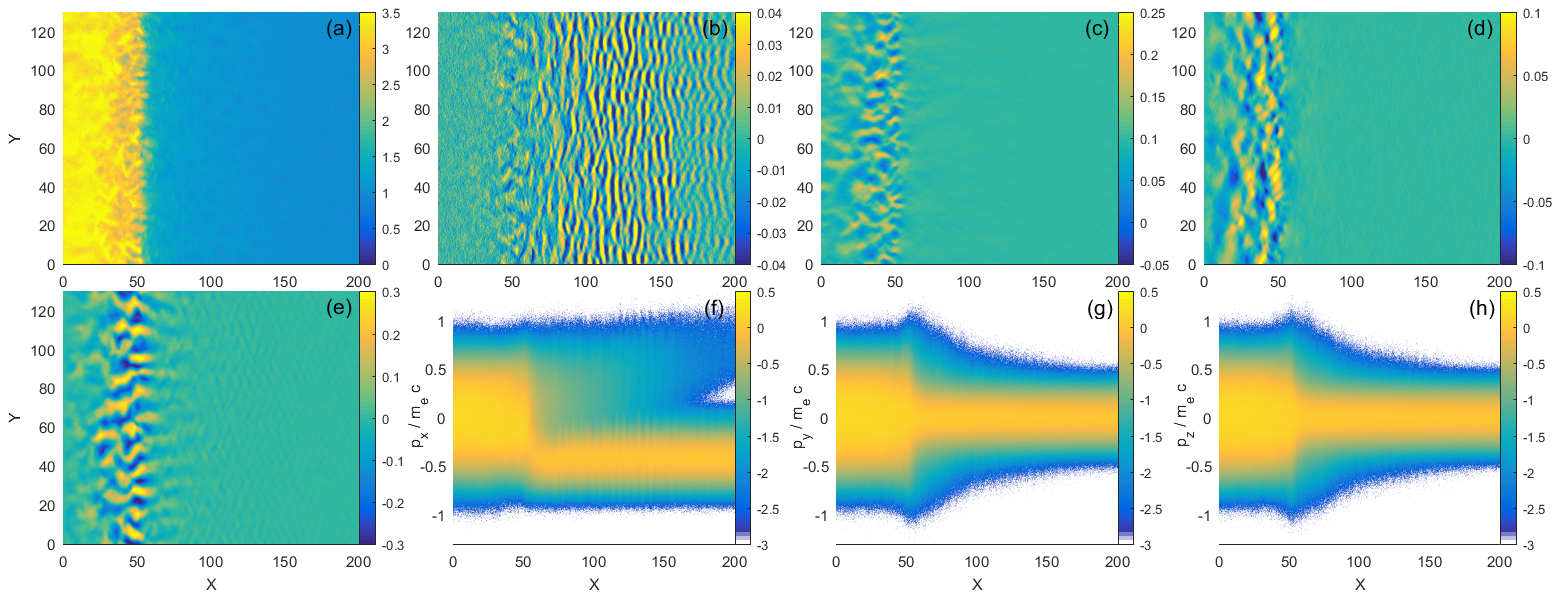}
\caption{The plasma and field distributions at the time $t=380$: Panel (a) shows the plasma density. The distribution of the electric $E_x$ component is displayed in (b). The magnetic $B_x$, $B_y$ and $B_z$ components are depicted by panels (c-e), respectively. The electron phase space density distributions $f_e(x,p_x)$, $f_e(x,p_y)$ and $f_e(x,p_z)$, which have been integrated over $y$, are shown in (f-h), respectively. They are normalized to the peak density in the upstream region and the color scale is 10-logarithmic.}
\label{figure6}
\end{figure*}

\begin{figure*}
\includegraphics[width=\textwidth]{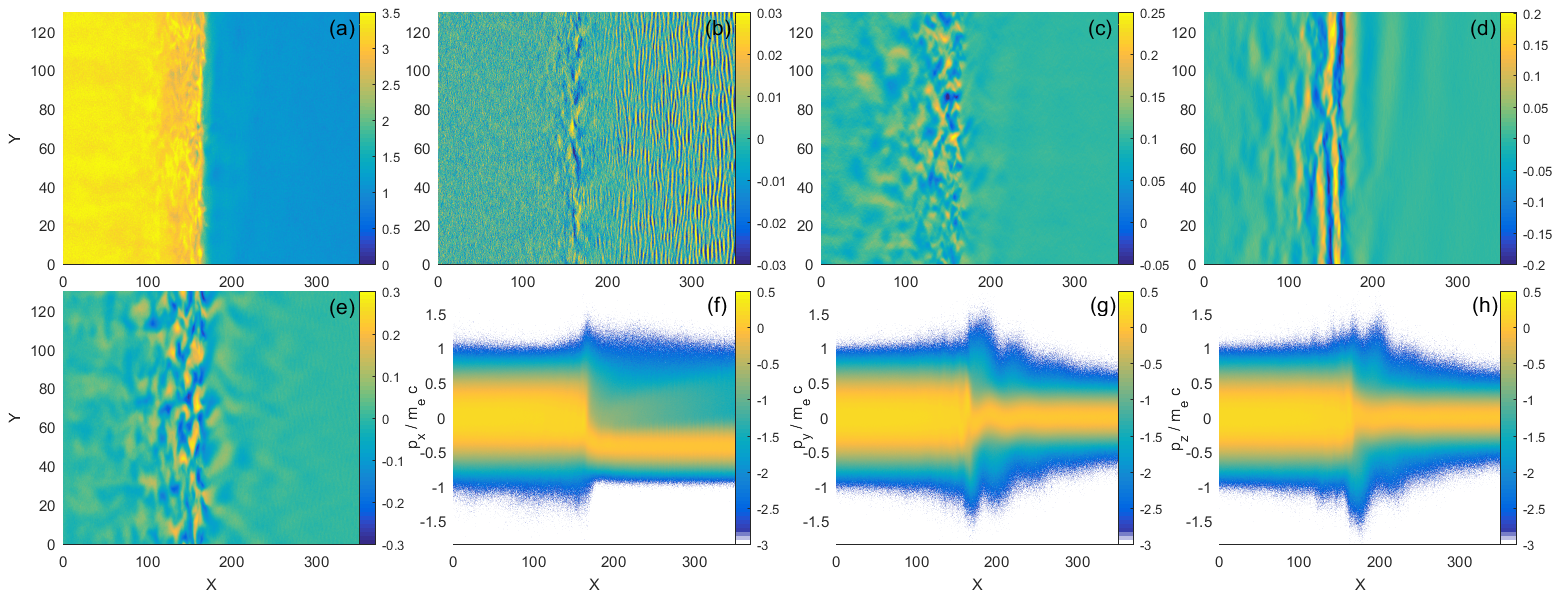}
\caption{The plasma and field distributions at the time $t=1100$: Panel (a) shows the plasma density. The distribution of the electric $E_x$ component is displayed in (b). The magnetic $B_x$, $B_y$ and $B_z$ components are depicted by panels (c-e), respectively. The electron phase space density distributions $f_e(x,p_x)$, $f_e(x,p_y)$ and $f_e(x,p_z)$, which have been integrated over $y$, are shown in (f-h), respectively. They are normalized to the peak density in the upstream region and the color scale is 10-logarithmic.}
\label{figure7}
\end{figure*}

\begin{figure*}
\includegraphics[width=\textwidth]{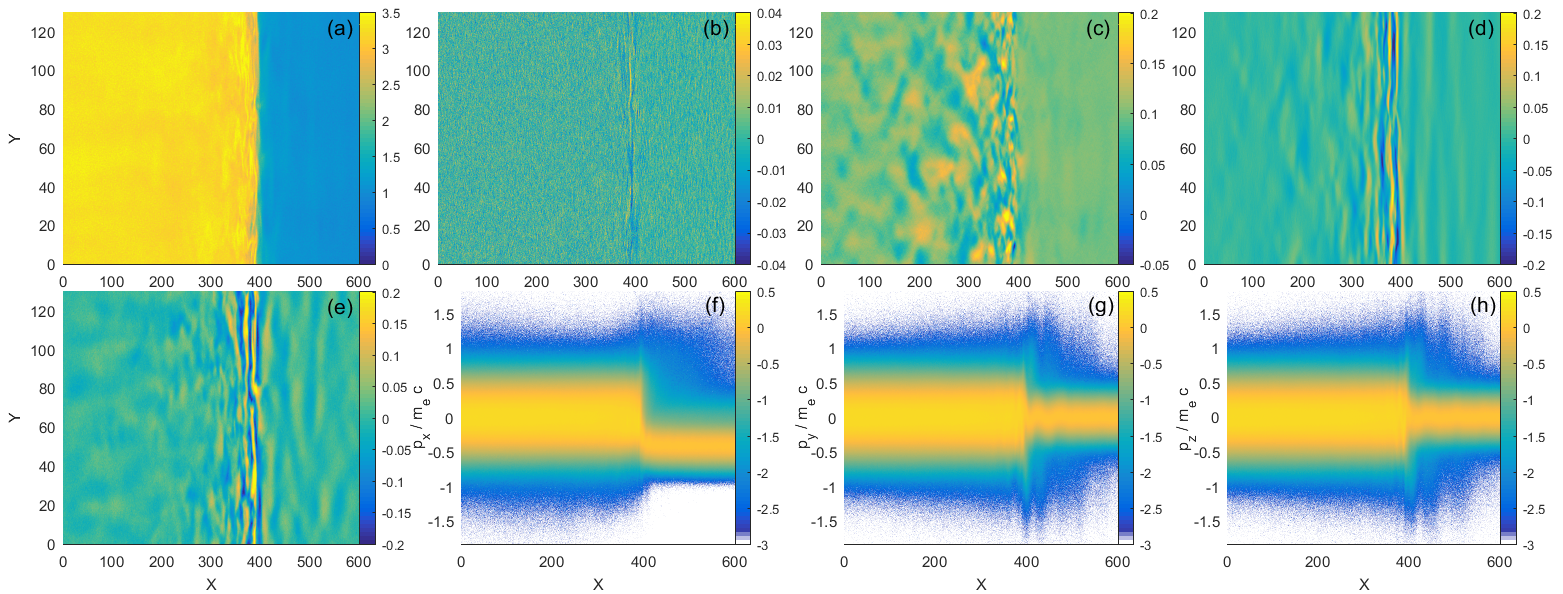}
\caption{The plasma and field distributions at the time $t=2500$: Panel (a) shows the plasma density. The distribution of the electric $E_x$ component is displayed in (b). The magnetic $B_x$, $B_y$ and $B_z$ components are depicted by panels (c-e), respectively. The electron phase space density distributions $f_e(x,p_x)$, $f_e(x,p_y)$ and $f_e(x,p_z)$, which have been integrated over $y$, are shown in (f-h), respectively. They are normalized to the peak density in the upstream region and the color scale is 10-logarithmic.}
\label{figure8}
\end{figure*}

We consider first the time evolution of the total plasma density, that of the field energy density $P_E = \epsilon_0\langle (E_x^2+E_y^2)\rangle_y/(2n_0k_BT_0)$ of the in-plane electric field and that of the field energy densities $P_{Bx}=\langle B_x^2 \rangle_y/(2\mu_0n_0k_BT_0)$, $P_{By}=\langle B_y^2 \rangle_y /(2\mu_0n_0k_BT_0)$ and $P_{Bz}=\langle B_z^2 \rangle_y/(2\mu_0n_0k_BT_0)$. We integrate these quantities over $y$ as indicated by the subscript of the brackets. Figure~\ref{figure5} shows the box-averaged plasma density and the square roots of the field energy densities.
Furthermore we give in Figs.~\ref{figure6}-\ref{figure8} snapshots of the plasma at three representative times; early time before pair-Alfvén growth ($t=380$), when the pair-Alfvén mode has saturated
($t=1100$), and late time when transient effects have ended ($t=2500$).

Figure~\ref{figure5}(a) shows that a shock front rapidly forms close to the wall
in the 2D simulation. We find already at $t\approx 100$ a boundary that
separates the downstream plasma with mean density $\approx 3.5$ from the
upstream plasma. This shock front propagates at the speed $v_{f2}=0.16c$ in the
positive $x$ direction, approximately $4/3$ the speed of the 1D shock. This
difference can be accounted for by the increased efficiency of particle-wave
scattering arising from the filamentation-driven turbulence in the 2D case. 
This efficiency lets the shock establish itself quickly and its propagation speed is set by
the pressure of a thermal downstream plasma. The shock forms later in the 1D simulation and
its downstream region is initially not in a thermal equilibrium. We measure the shock speed $v_{f2}$ in the box frame. Its speed in the rest frame of the upstream plasma is $v_{f2}+|v_b|=0.58c$, which exceeds the pair-Alfv\'en speed $v_A$ by the factor 6. 

A dilute plasma beam outruns the shock in Fig. \ref{figure5}(a) and reaches an $x$-position of 700 at $t\approx
1700$, which amounts to the speed $0.4c$. This speed matches the mean speed
modulus of the upstream plasma. The front of the 
dilute beam is initially trailed by a second beam with the density $\approx
1.3$, which reaches $x\approx 300$ at $t\approx 1300$ and is absorbed by the
shock at later times. Figure \ref{figure5}(b) shows a broad electrostatic pulse
that outruns the shock that is centered on the front of the dilute plasma beam.
Figures \ref{figure5}(c-e) show peaks in their energy density that are
positioned at the shock. The energy density of the $B_x$ component does not
extend far beyond the shock position. The energy densities of the $B_y$ and
$B_z$ components reach further into the upstream region than that of $B_x$, and
the energy density of the $B_z$ component expands even faster than that of $B_y$.

We can understand the nature of these structures with the help of the spatial distributions of the electromagnetic fields and the phase space density distributions of the electrons and positrons. 
Figure \ref{figure6} shows these at time $t=380$. The density jump caused by the shock is located at $x=50$. A downstream region with the
density $3.5$ has formed in Fig. \ref{figure6}(a), which is separated by a
sharp shock boundary from the inflowing upstream region.  Figure
\ref{figure6}(b) reveals strong electrostatic waves just ahead of the shock. On
average, their wave vector is aligned with the x-direction, which means that
these structures can be resolved by a 1D simulation. We thus identify them with
the phase space vortices we found in Fig. \ref{figure3}(a) in the interval $150
\le x \le 250$. The width of these phase space vortices was on the order of few
plasma skin depths, which matches the wavelength of the waves in Fig.
\ref{figure6}(b). The mean speed of the vortices is larger than the shock speed,
explaining why they outrun the shock in Fig. \ref{figure5}(a). The fastest phase space vortices involve only a very dilute plasma, which is not resolved by the color scale in Fig. \ref{figure5}(a). Hence the electrostatic waves seem to outrun the front of the dilute plasma. 

Figures \ref{figure6}(c-e) show the spatial distributions of the magnetic field
components. The strongest component is $B_z$. Initally the filamentation
sets up a periodic oscillation of plasma density along $y$. This then drives turbulence with $\mathbf B
\parallel z$. This asymmetry between the $y$ and $z$ directions is, however, an artefact of the dimensionality of
the simulation, in full 3D the filamentation instability would similarly drive magnetic turbulence along $y$.
The filamentation-driven turbulence initially dominates since its growth rate is proportional to $\omega_p$ whereas the growth rate of the
pair-Alfvén mode is proportional to the electron cyclotron frequency $\omega_c$
which is smaller by a factor of 10. Hence the fastest growth is seen in $B_z$ which
saturates near the shock front almost instantly. Figures \ref{figure5}(c, d) show that the peak of $B_z$ energy density is colocated with that of $B_x$ but not with that
of $B_y$. The modulation of $B_x$ is thus a consequence of the modulation of
$B_z$ and presumably needed to fulfill $\nabla \cdot \mathbf{B}=0$. The magnetic
$B_y$ component shows oscillations in the downstream region of the shock, which
suggests that pair-Alfv\'en waves have also grown in the 2D simulation. 
Their amplitude is barely a third of that of the filamentation modes. At first glance, one may thus conclude that the filamentation modes mediate the shock at this time. However, resonant interactions between particles and the pair-Alfv\'en wave could enhance their scattering. It has been reported that such resonant interactions can play an important role in the transition layer of Alfv\'enic shocks in electron-ion plasma\cite{Zekovic2018,Zekovic2019} and the same may hold in our simulation. We also have to take into account that the magnetic amplitude of the pair-Alfv\'en wave is comparable to $B_0=0.1$. The wave is thus already in a non-linear regime. 

Figure \ref{figure6}(f) reveals a bulk distribution of the plasma centered on
$p_x\approx -0.5m_ec$ for $x>50$; this is the upstream plasma. The mean velocity
jumps to $p_x=0$ at the location $x=50$ of the shock. A larger momentum spread
of the particles downstream of their shock implies that the plasma has been
heated by crossing the shock. A dilute hot plasma beam with $p_x>0$ outruns the
shock. The phase space vortices in this beam cannot be seen due to the phase
space density distribution being integrated over $y$. The beam also drives the
electrostatic waves in Fig. \ref{figure6}(b),  a two-stream instability. Such
waves typically have speed comparable to that of the beam with the lower plasma
frequency (in this case the dilute beam reflected from the wall). Here the waves
have velocity $\approx 0.27c$ in the simulation frame, and so quickly outrun the shock.

Figures \ref{figure6}(g, h) demonstrate
that the plasma has been heated by the shock passage also along the other
directions. Both momentum distributions are evidence of energetic particles
ahead of the shock. Their peak momentum decreases with distance from the shock
and the distribution of the energetic particles joins with the distribution of the upstream plasma at $x=200$. This implies that the energetic beam at $x>50$ and $p_x>0$ in Fig. \ref{figure6}(f) consists of particles with relatively low values of $p_y$ and $p_z$. The beam is thus fed by energetic downstream particles with a momentum vector that is almost aligned with $\mathbf{B}_0$.  

Figure \ref{figure7} shows the same distributions at time $t=1100$.
The plasma density distribution in Fig. \ref{figure7}(a) reveals a shock at
$x\approx 160$. Its transition layer has not broadened along $x$ and it still
compresses the plasma to a downstream density $\approx 3.5$. The electrostatic
waves in Fig. \ref{figure7}(b) have propagated ahead of the shock as already
demonstrated by Fig. \ref{figure5}(b). These waves apparently form only at early
times. The two-stream instability grows quickly if two dense beams interact
while remaining well-separated along the velocity direction. This is the case
here prior to the formation of the shock because the incoming plasma and the
plasma reflected by the wall can interact. We find two well separated beams in
Fig. \ref{figure6}(f) for $x>180$. Once the shock has formed, the plasma that
makes it upstream is hotter and less dense, and we no longer have a bimodal
distribution in $p_x$, hence the two-stream instability is quenched. 

The amplitudes of $B_y$ and $B_z$ are now comparable. The $B_z$ component shows
small patches, which oscillate rapidly along $y$, and are correlated with the
oscillations of $B_x$. The wave vector of the oscillations of $B_y$ is practically aligned with $x$;
the magnetic field oscillations are thus tied to pair-Alfv\'en modes. We find
oscillations of $B_y$ with the same orientation to the left and right of the
shock at $x=160$. We observe waves to both sides of the shock also in Fig.
\ref{figure7}(e). They appear more turbulent. The oscillations of $B_y$ and
$B_z$ appear identical to those in the 1D simulation. We thus infer from their
different distributions in Fig. \ref{figure7} that the oscillations of $B_y$ are
caused by pair-Alfv\'en waves while those of $B_z$ are tied to filamentation
modes and pair-Alfv\'en modes. The growth rate of the filamentation instability
is greatest for particles streaming at relativistic speeds. These modes can then
be driven even by dilute relativistic beams. This explains why we find the growth of waves in Fig. \ref{figure5}(e) in the $B_z$ distribution well ahead of the $x-$interval in which the pair-Alfv\'en modes grow in Fig. \ref{figure5}(d).  

Figure \ref{figure7}(f) shows that the particles escaping upstream of the shock no longer form a beam that could drive electrostatic instabilities. The existing waves propagate ahead of the shock. The momentum distributions along $p_y$ and $p_z$ show oscillations ahead of the shock. Upstream particles are deflected by the magneto-waves close to and ahead of the shock. The large oscillation amplitude implies that the particle speed is large relative to the wave speed. This suggests in turn that the speed of the pair-Alfv\'en waves changes as they approach the shock. The piled-up waves form a shock precursor.

Figure \ref{figure8} shows the equivalent spatial distributions at the time $t=2500$.
The shock is now located at $x=400$ in Fig. \ref{figure8}(a) and the uniform density of the downstream plasma suggests that it has been fully thermalized. The electrostatic waves have moved far upstream at this time and are no longer captured by the resolved x-interval in Fig. \ref{figure8}(b). Electric field oscillations are seen close to the shock boundary; they are not necessarily electrostatic since we find time-varying magnetic fields that can induce electric fields via Faraday's law. The magnetic $B_y$ and $B_z$ components show oscillations with a wave vector that is aligned with $x$. This suggests that pair-Alfv\'en waves are now mediating the shock. Figures \ref{figure8}(f-h) show a thermalized downstream region that is separated by the shock from the upstream plasma. The mean velocity along $x$ of the plasma changes drastically across the shock and so does the plasma temperature. The inflowing upstream particles still gyrate in the magnetic field of the pair-Alfv\'en modes. 

\begin{figure*}
\includegraphics[width=\textwidth]{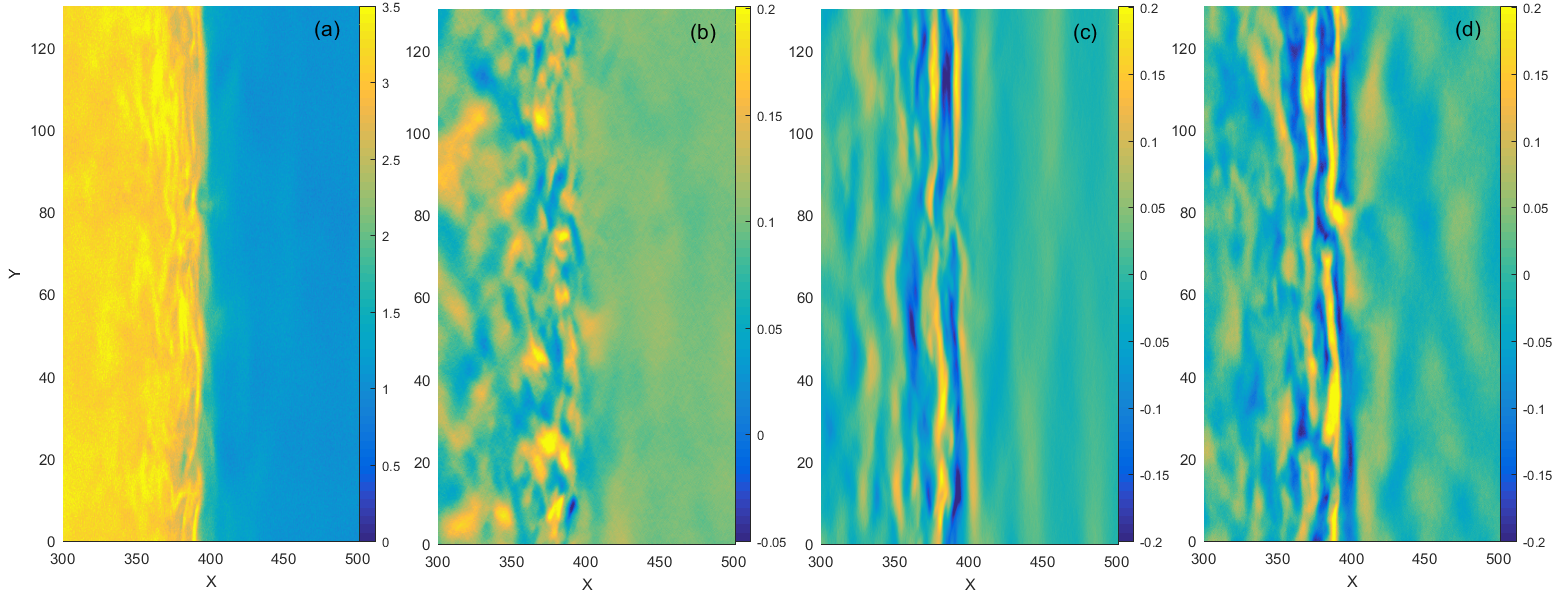}
\caption{As in \ref{figure7} this figure shows the plasma and field
  distributions at the simulation end time $t=t_{\mathrm{sim}}=2500$, but zoomed
  in around $x=400$ to show detail at the shock front. Panel (a) shows the plasma density. The magnetic $B_x$, $B_y$ and $B_z$ components are depicted by panels (b-d), respectively.}
\label{figure9}
\end{figure*}

Figure \ref{figure9} zooms in on the shock front.
The shock front in Fig. \ref{figure9}(a) is sharp even on this spatial scale. The $B_x$ component shows rapid oscillations along $y$, which suggests that the filamentation instability has not completely disappeared. Figures \ref{figure9}(c, d) demonstrate that waves at the shock front are quasi-planar. The $B_z$ component shows oscillations along $y$ ahead of the shock and also at the shock.

How can we explain that initially the filamentation instability dominated while we observe predominantly pair-Alfv\'en modes at later times? The fact that we observed the filamentation instability at early times indicates that its waves have a larger linear growth rate meaning that they grow faster from noise levels to their nonlinear saturation\cite{Bret2014}. The filamentation modes thermalized the pair plasma before the pair-Alfv\'en wave instability could grow. The growth of waves upstream of the shock, which are convected to the shock, implies that instabilities at the shock no longer grow from noise levels. Oscillations in the $B_y$ component can only be caused by pair-Alfv\'en waves. The upstream waves provide a seed for the instability at the shock, which causes the growth of strong pair-Alfv\'en modes at the shock. The upstream waves in the $B_z$ distribution in Fig. \ref{figure9}(d) are composed of filamentation modes and pair-Alfv\'en modes, which implies that both instabilities at the shock are boosted by these seed perturbations.  

\begin{figure}
\includegraphics[width=\columnwidth]{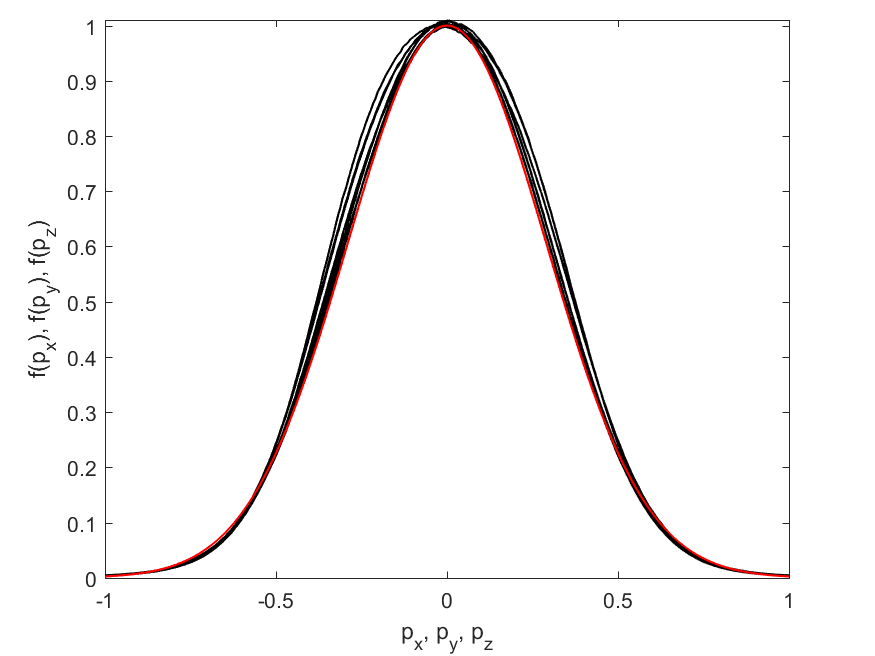}
\caption{Electron momentum distributions at $t=t_{sim}$: The electron phase space distributions $f_e(x,p_x)$, $f_e(x,p_y)$ and $f_e(x,p_z)$ have been integrated over the intervals $0\le x \le 130$, $130\le x \le 260$ and $260 \le x \le 390$. The resulting 9 curves are plotted together with a Maxwellian distribution with the temperature 43 keV.}
\label{figure10}
\end{figure}

Finally we want to test how well the plasma has been thermalized by the shock crossing and what temperature it reached. We subdivide the downstream regions in Fig. \ref{figure8}(f-g) into the intervals $0\le x \le 130$, $130 \le x \le 260$ and $260 \le x \le 390$ and integrate them separately over $x$. We integrate the three momentum distributions separately and obtain a total of 9 momentum distributions. We plot in Fig. \ref{figure10} all distributions into the same panel.
All distributions agree well and can be approximated by a Maxwellian with the
temperature 43 keV. We find no thermal anisotropies and no variations of the
temperature with $x$; the downstream plasma has thus been well thermalized by
the shock crossing.



\section{Discussion}

In summary, we have demonstrated the existence of a novel type of collisionless shock which may occur in 
mildly-relativistic pair plasmas, such as those thought to exist in the inner regions of jets that are emitted by 
microquasars. These shocks are initially fed by turbulence generated by the filamentation instability 
as it was found in previous PIC simulations of unmagnetized leptonic shocks\cite{Kazimura1998,Spitkovsky2005,Chang2008,Nishikawa2009,Sironi2009}. We aligned a magnetic field 
with the average shock normal. The magnetic field amplitude was not large enough to suppress the filamentation instability as in the simulation by Hededal et al.\cite{Hededal2005}. Instead it provided a new unstable wave branch, namely the pair-Alfv\'en 
wave. We compared the results of a 1D PIC simulation study, where the filamentation instability was 
suppressed due to the alignment of the simulation box with the magnetic field, with that of a 2D
study. 

Initially, both simulations provided different results. The instability that mixed the inflowing pair plasma with
the one reflected by the simulation box boundary was an electrostatic two-stream instability in the 1D case and
the filamentation instability in the 2D simulation. Electrostatic waves with an electric field pointing
along the magnetic field can mix particles only along the magnetic field, while particles can be deflected 
in all directions by the magnetic filamentation modes together with the background magnetic field. The plasma
downstream of the shock was thus far from a thermal equilibrium in the 1D simulation while it
was almost thermal in the 2D one. Eventually, a pair-Alfv\'en wave instability thermalized the downstream
plasma also in the 1D simulation.

Energetic downstream particles were able to outrun the forming shock before it settled into its final state. The particles
formed a beam that drove electrostatic instabilities upstream of the forming shock. These fast waves outran
the shock and they damped out eventually. Once the shock was established, the escaping particles formed a
diffuse energetic upstream population that was no longer able to drive electrostatic waves. The leaking downstream particles increased the mean thermal energy of the upstream plasma in the direction of $\mathbf{B}_0$ without introducing separate particle beams. Such a distribution is close to a bi-Maxwellian. The pair-Alfv\'en wave instability that developed ahead of the shock may thus be similar to the firehose instability discussed by Schlickeiser~\cite{Schlickeiser2010}. The pair-Alfv\'en waves it seeded were slower than the electrostatic ones and the shock could catch up with them. Their pile-up changed the shock into an Alfv\'enic one in the 1D simulation. Pair-Alfv\'en modes coexisted with filamentation modes in the transition layer of the two-dimensional 
shock. Both modes had comparable amplitudes of the out-of-plane magnetic field. Filamentation modes with in-plane 
magnetic fields are excluded in a 2D geometry and the in-plane magnetic field was tied exclusively to 
pair-Alfv\'en modes. We note in this context that the pair-Alfv\'en wave in a pair plasma with identical distributions of electrons and positrons has a linear polarization. Pair-Alfv\'en waves with an in-plane magnetic field may thus be decoupled from waves with a magnetic field that points out of the simulation plane. Future 3D PIC simulation studies have to test if the pair-Alfv\'en wave can replace the filamentation mode in a realistic geometry. Future work will also have to determine how the shock structure changes with the amplitude and direction of the background magnetic field. Finally one also has to test if pair-Alfv\'en waves keep their linear polarization in the non-linear regime.

Since potential sites for these shocks are ubiquitous in the high-energy sky we expect that the radiative signal 
produced by the accelerated leptons should be observable, at least in regimes with low optical depth to the base 
of the jet, e.g. radio. To achieve this, further work should examine more closely the expected radiation signature 
and provide a prediction for observations.

Additionally we may expect to see significant acceleration for ions at these shocks \cite{Sironi2015}, and so it 
should be investigated if this can be shown to occur by adding a low number density ''cosmic-ray'' plasma species 
to future simulations. Furthermore the cosmic-ray instability and back-reaction of the accelerated ions on the turbulence 
may materially affect the shock structure. This was recently demonstrated for electron-ion shocks \cite{Crumley2019}
but not for electron-positron-ion shocks. We defer this to a future publication.

\textbf{Acknowledgements:} The simulation was performed on resources provided by the Swedish National Infrastructure for Computing (SNIC) at HPC2N (Ume\aa). AP acknowledges support from the EU via the ERC grant (O.M.J.). This research was supported in part by the National Science Foundation under Grant No. NSF PHY-1748958 and by the KITP at Santa Barbara. Raw data were generated at the HPC2N large scale facility.  Derived data supporting the findings of this study are available from the corresponding author upon reasonable request.


\begin{thebibliography}{100}
\bibitem{Margon1979} B. Margon, H. C. Ford, S. A. Grandi, and R. P. S. Stone, Astrophys. J. \textbf{233}, L63 (1979).
\bibitem{Falcke1996} H. Falcke, and P. L. Biermann, Astron. Astrophys. \textbf{308}, 321 (1996).
\bibitem{Fender2004} R. P. Fender, T. M. Belloni, and E. Gallo, Mon. Not. R. Astron. Soc. \textbf{355}, 1105 (2004).
\bibitem{Bromberg2011} O. Bromberg, E. Nakar, T. Piran, and R. Sari, Astrophys. J. \textbf{740}, 100 (2011).
\bibitem{Dieckmann2019} M. E. Dieckmann, D. Folini, I. Hotz, A. Nordman, P. Dell'Acqua, A. Ynnerman, and R. Walder, Astron. Astrophys. \textbf{621}, A142 (2019).
\bibitem{Kazimura1998} Y. Kazimura, J. I. Sakai, T. Neubert, and S. V. Bulanov, Astrophys. J. \textbf{498}, L183 (1998).
\bibitem{Spitkovsky2005} A. Spitkovsky, AIP Conf. Proc. \textbf{801}, 345 (2005).
\bibitem{Chang2008} P. Chang, A. Spitkovsky, and J. Arons, Astrophys. J. \textbf{674}, 378 (2008).
\bibitem{Nishikawa2009} K. I. Nishikawa, J. Niemiec, P. E. Hardee, M. Medvedev, H. Sol, Y. Mizuno, B. Zhang, M. Pohl, M. Oka, and D. H. Hartmann, Astrophys. J. \textbf{698}, L10 (2009).
\bibitem{Sironi2009} L. Sironi, and A. Spitkovsky, Astrophys. J. \textbf{698}, 1523 (2009).
\bibitem{Marcowith2016} A. Marcowith, A. Bret, A. Bykov, M. E. Dieckman, L. O. Drury, B. Lembege, M. Lemoine, G. Morlino, G. Murphy, G. Pelletier, I. Plotnikov, B. Reville, M. Riquelme, L. Sironi, and A. S. Novo, Rep. Prog. Phys. \textbf{79}, 046901 (2016). 
\bibitem{Dieckmann2017} M. E. Dieckmann, and A. Bret, J. Plasma Phys. \textbf{83}, 905830104 (2017).
\bibitem{Weibel1959} E. Weibel, Phys. Rev. Lett. \textbf{2}, 83 (1959).
\bibitem{Dieckmann2018MNRAS} M. E. Dieckmann, and A. Bret, Mon. Not. R. Astron. Soc. \textbf{473}, 198 (2018).
\bibitem{Bret2010} A. Bret, L. Gremillet, and M. E. Dieckmann, Phys. Plasmas \textbf{17}, 120501 (2010).
\bibitem{Hededal2005} C. B. Hededal, and K. I. Nishikawa, Astrophys. J \textbf{623}, L89 (2005).
\bibitem{Bret2006} A. Bret, M. E. Dieckmann, and C. Deutsch, Phys. Plasmas \textbf{13}, 082109 (2006).
\bibitem{Schlickeiser2010} R. Schlickeiser, Open Plasma Phys. J. \textbf{3}, 1 (2015).
\bibitem{Bret2018} A. Bret, and R. Narayan, J. Plasma Phys. \textbf{84}, 905840604 (2018).
\bibitem{DieckmannPPCF2019} M. E. Dieckmann, D. Folini, A. Bret, and R. Walder, Plasma Phys. Controll. Fusion \textbf{61}, 085027 (2019). 
\bibitem{Bell2004} A. R. Bell, Mon. Not. R. Astron. Soc. \textbf{353}, 550 (2004).
\bibitem{Park2015} J. Park, D. Caprioli, and A. Spitkovsky, Phys. Rev. Lett. \textbf{114}, 085003 (2015).
\bibitem{Caprioli2018} D. Caprioli, H. Zhang, and A. Spitkovsky, J. Plasma Phys. \textbf{84}, 715840301 (2018). 
\bibitem{Keppens2019} R. Keppens, and H. Goedbloed, J. Plasma Phys. \textbf{85}, 175850101 (2019).
\bibitem{Crumley2019} P. Crumley, D. Caprioli, S. Markoff, and A. Spitkovsky, Mon. Not. R. Astron. Soc. \textbf{485}, 5105 (2019).
\bibitem{Arber2015} T. D. Arber, K. Bennet, C. S. Brady, A. Lawrence-Douglas, M. G. Ramsay, N. J. Sircombe, P. Gillies, R. G. Evans, H. Schmitz, A. R. Bell, and C. P. Ridgers, Plasma Phys. Controll. Fusion \textbf{57}, 113001 (2015). 
\bibitem{Dieckmann2004} M. E. Dieckmann, A. Ynnerman, S. C. Chapman, G. Rowlands, and N. Andersson, Phys. Scripta \textbf{69}, 456 (2004).
\bibitem{Jao2014} C. S. Jao, and L. N. Hau, Phys. Rev. E \textbf{89}, 053104 (2014).
\bibitem{Morse69} R. L. Morse, and C. W. Nielsen, Phys. Rev. Lett. \textbf{23}, 1087 (1969).
\bibitem{Roberts67} K. V. Roberts and H. L. Berk, Phys. Rev. Lett. \textbf{19}, 297 (1967). 
\bibitem{Gary2009} P. S. Gary, and H. Karimabadi, Phys. Plasmas \textbf{16}, 042104 (2009). 
\bibitem{Stewart1992} G. A. Stewart and E. W. Laing, J. Plasma Phys. \textbf{47}, 295 (1992).
\bibitem{Zekovic2018} V. Zekovic, and B. Arbutina, Nucl. Part. Phys. Proc. \textbf{297}, 53 (2018).
\bibitem{Zekovic2019} V. Zekovic, Phys. Plasmas \textbf{26}, 032106 (2019). 
\bibitem{Bret2014} A.~Bret, A.~Stockem, R.~Narayan, and L.~O.~Silva, Phys. Plasmas \textbf{21}, 072301 (2014).
\bibitem{Sironi2015} L. Sironi, U. Keshet, and M. Lemoine, Space Sci. Rev. \textbf{519}, 191 (2015).

 
\end{thebibliography}
\end{document}